%% file: cosmicmr.tex
\documentclass[12pt]{article}
\usepackage{graphicx}

\newcommand\pubnumber{DPF2015-171}
\newcommand\pubdate{\today}

\def\napoli{Department of Physics and Astronomy\\
University of South Carolina\\
 Columbia, SC 29205 United States}

\def\Title#1{\begin{center} {\Large #1 } \end{center}}
\def\Author#1{\begin{center}{ \sc #1} \end{center}}
\def\Address#1{\begin{center}{ \it #1} \end{center}}

\newcommand\pubblock{\rightline{\begin{tabular}{l} \pubnumber\\
         \pubdate  \end{tabular}}}
\newenvironment{Abstract}{\begin{quotation}  }{\end{quotation}}
\newenvironment{Presented}{\begin{quotation} \begin{center} 
             PRESENTED AT\end{center}\bigskip 
      \begin{center}\begin{large}}{\end{large}\end{center} \end{quotation}}

\input econfmacros.tex

\begin{document}
\begin{titlepage}
\pubblock

\vfill
\Title{Cosmic Ray Induced EM Showers in the NO$\nu$A Detectors}
\vfill
\Author{ Hongyue Duyang \\ For the NO$\nu$A Collaboration}
\Address{\napoli}
\vfill
\begin{Abstract}
The NO$\nu$A experiment is an electron neutrino appearance neutrino oscillation experiment at Fermilab. Electron neutrino events are identified by the electromagnetic (EM) showers induced by electrons in the final state of neutrino interactions. EM showers induced by cosmic muons or rock muons, are abundant in NO$\nu$A detectors. We use a Muon-Removal Technique to get pure EM shower samples from cosmic and rock muon data. Those samples can be used to characterize the EM signature and provide valuable checks of the MC simulation, reconstruction, PID algorithms, and calibration across the NO$\nu$A detectors.
\end{Abstract}
\vfill
\begin{Presented}
DPF 2015\\
The Meeting of the American Physical Society\\
Division of Particles and Fields\\
Ann Arbor, Michigan, August 4--8, 2015\\
\end{Presented}
\vfill
\end{titlepage}
\def\thefootnote{\fnsymbol{footnote}}
\setcounter{footnote}{0}

\section{Introduction}

NO$\nu$A (NuMI Off-Axis $ \nu_{e} $ Appearance) is a long-baseline $\nu_{\mu}$ to $\nu_e$ neutrino oscillation experiment designed to determine the neutrino mass hierarchy and constrain CP-violation \cite{novadesign}. It uses neutrino beam generated by the Main Injector (NuMI) at Fermilab \cite{numi}, and measures the neutrino flavors by the near detector (ND, 0.3 kiloton) located at Fermilab, and the far detector (FD, 14 kiloton) at Ash River, MN. 
The NO$\nu$A detectors are constructed from liquid scintillator contained inside extruded PVC modules. 
The extrusions are assembled in alternating layers of vertical and horizontal planes, which are 0.15 radiation lengths in width, optimized for electron reconstruction. 

NO$\nu$A finds signal $ \nu_{e} $ events by detecting electrons in the final state of charged current electron neutrino interaction ($ \nu_{e} $-CC) \cite{novafa}. This requires correct modeling, reconstruction and particle identification (PID) algorithms for the electromagnetic (EM) showers. 
Calibration effects such as attenuation and alignment should also be under control. 

Cosmic ray muons are abundant (148 kHz) in the NO$\nu$A far detector which is on the surface. In the near detector, rock muons are abundant. 
They induce EM showers by three different means: energetic muons undergoing bremsstrahlung radiation (Brem), muons decaying into electrons in flight (DiF), and muons stoping in the detectors and decaying into Michel electrons. Michel electrons' energy is small compared to $ \nu_{e} $ events (0.5GeV$ \sim$ 4GeV), and have been used for calibration. Brem and DiF, on the other hand, provide abundant EM showers at few-GeV energy region. This sample makes possible s data-driven method to benchmark EM shower modeling, PID, as well as the detector calibration.

\begin{figure}
\begin{center}
\includegraphics[width=0.7\textwidth]{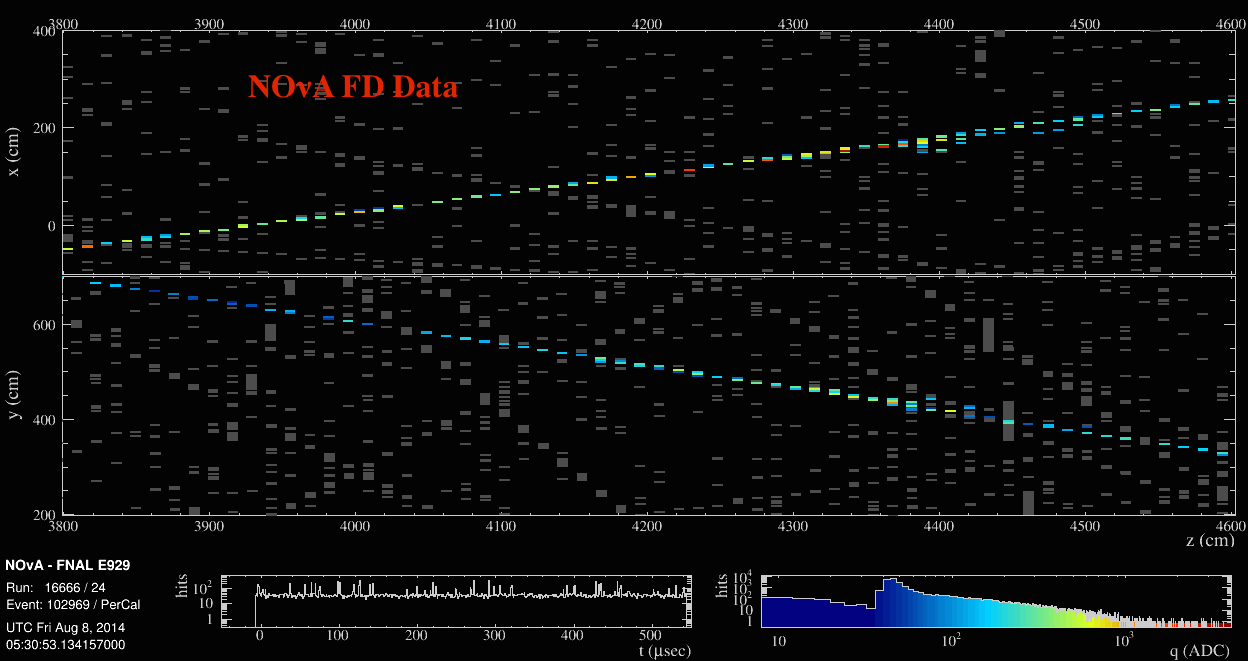}\\
\includegraphics[width=0.7\textwidth]{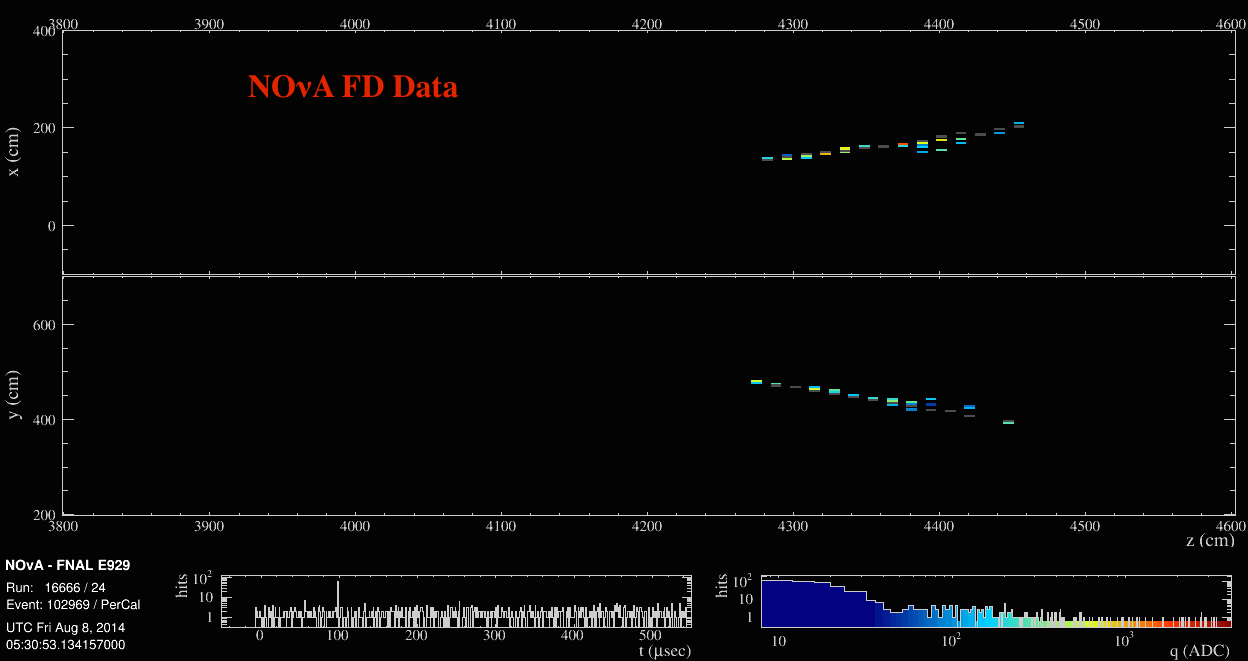}
\end{center}
\caption{Top: event display cosmic muon candidate with Electromagnetic (EM) Bremsstrahlung (Brem) Shower from NO$\nu$A simulation. Bottom: event display of hits of the EM shower after the removal of hits associated with the muon track from NO$\nu$A simulation. What left are the hits of Brem Showers. \label{fdevd}}
\end{figure}

For this purpose, a cosmic muon-removal (MR) algorithm is developed to identify the Brem and DiF showers, remove the muons, and save the remaining EM showers at raw digit level. Figure \ref{fdevd} show examples in the FD for EM shower events before and after MR.
The shower digits can then be put into standard $\nu_e$ reconstruction and PID algorithms. Data and MC comparison is performed with reconstructed shower variables and PID outputs to validate EM shower modeling and PID. Calibration effects can be checked by comparing PID efficiencies as function of vertex position.

\section{Procedure}

We take cosmic data and MC simulation in NO$\nu$A FD. Both are subject to standard cosmic reconstructions. The following procedure breaks into 4 steps. 

\textbf{1. Track Selection}: Apply selection cuts to search for cosmic track candidates in the far detector. Basically  We want the tracks to be through-going muons which are not too steep compared to the beam direction (cos$\theta>0.5$) and long enough to have a good chance of generating Brem showers (number of track planes $>$ 30).  Then to compare with beam $\nu_e$ events we want the shower to have energy close to the $\nu_e$ energy region ($0.5\sim5$ GeV). 

\textbf{2. Shower Finding}: Taking candidate tracks as input, an EM shower finding algorithm is used to identify the muons with possible EM shower dE/dx information. Muons deposit energy as a minimum ionization particle (MIP) in the detector cells. The additional EM shower hits can deposit much more energy in a small detector region and overlaps with muon energy. It is then possible to take the energy deposition of reconstructed cosmic tracks and look for the region where the energy is significantly greater than a MIP. 
More specifically,
If we find 5 consecutive planes with energy greater than 2 MIP, call it the shower start point;
If we find 5 consecutive plane with energy in the range of 0.5 MIP to 1.5 MIP, call it the shower end point;
If  we find both the shower start and end point, then the shower is identified as Brem; If we find only the shower start point, and reach the end of track without an end point, the shower is identified as an electron from DiF.  
Showers with energy deposition passing the energy cut are saved as raw digit objects.

\textbf{3. Muon Removal (MR)}: A Muon-Remove algorithm for EM showers is modified from the Muon-Removal algorithm for charged current events.\cite{mrcc}. 
It first looks at the slice where a cosmic EM shower is found.
In the case of DiF muon, with a shower region defined by the shower-finding algorithm, it just removes all hits outside that region and what left will be pure electron hits.
In the case of bremsstrahlung showers one additional problem is that we have a muon track inside the EM shower region. Therefore the muon-removal algorithm should remove hits that belong to muon track corresponding to the energy of a MIP in the shower region. 
All other hits in those slices where no shower is found are removed. Those hits not associated with any slices are also removed.

\section{EM Shower Reconstruction and Identification}

The saved shower raw digits with muon removed are put into the standard $\nu_e$ reconstruction. 
The reconstruction starts with clustering together hits from a single neutrino interaction in to a slice clustering hits by space-time coincidence[ref]. 
A modified Hough transform is used to identify prominent straight-line features in a slice. The slice vertex is defined by tuning the lines in an iterative procedure and finding the converge point.
Prongs are then reconstructed as groups of hits based on their distances to the lines associated with the reconstructed vertex [ref]. 

\begin{figure}
\begin{center}
\includegraphics[width=0.4\textwidth]{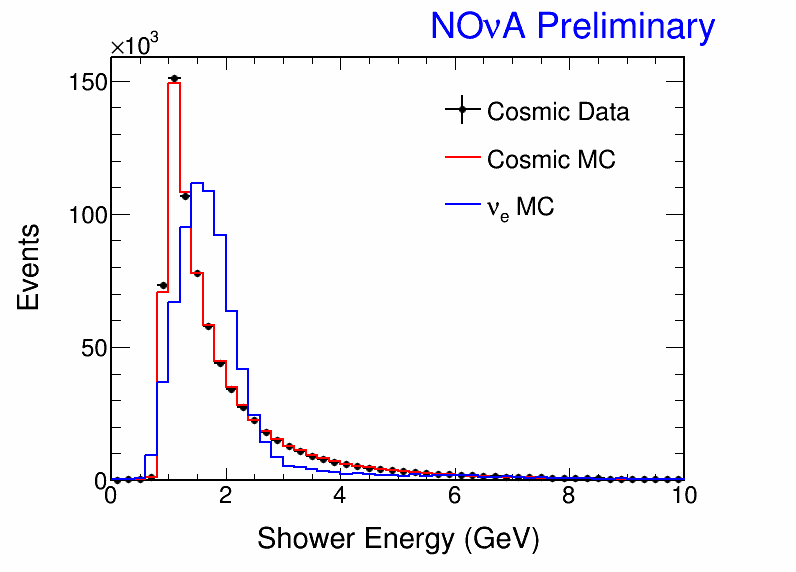}
\includegraphics[width=0.4\textwidth]{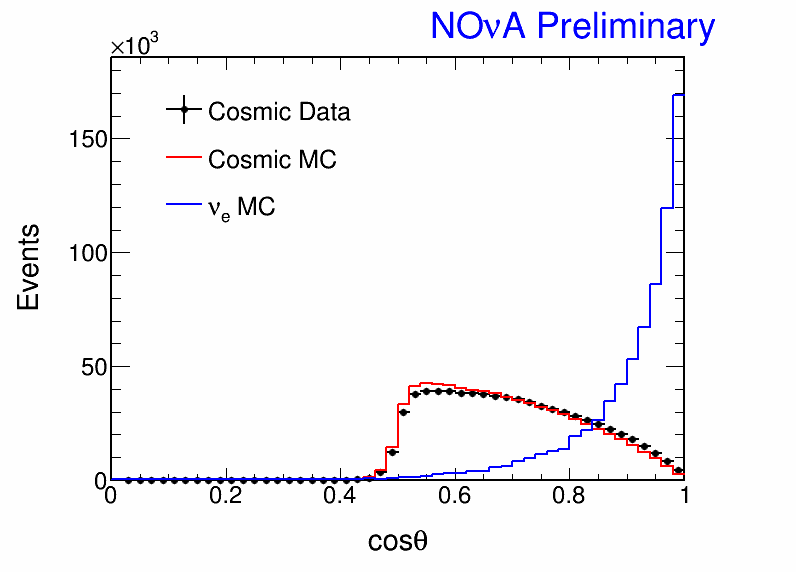}
\end{center}
\caption{Left: reconstructed shower energy after the cosmic Muon Removal(MR) in the far detector. 
Right: cosine of the reconstructed shower angle with respect to the beam direction. Distribution of $\nu_e$ events are drawn on the same plots. Cosmic EM showers show good data/MC agreement, and cover the range of $\nu_e$ events. \label{eshw}}
\end{figure}

\begin{figure}
\begin{center}
\includegraphics[width=0.4\textwidth]{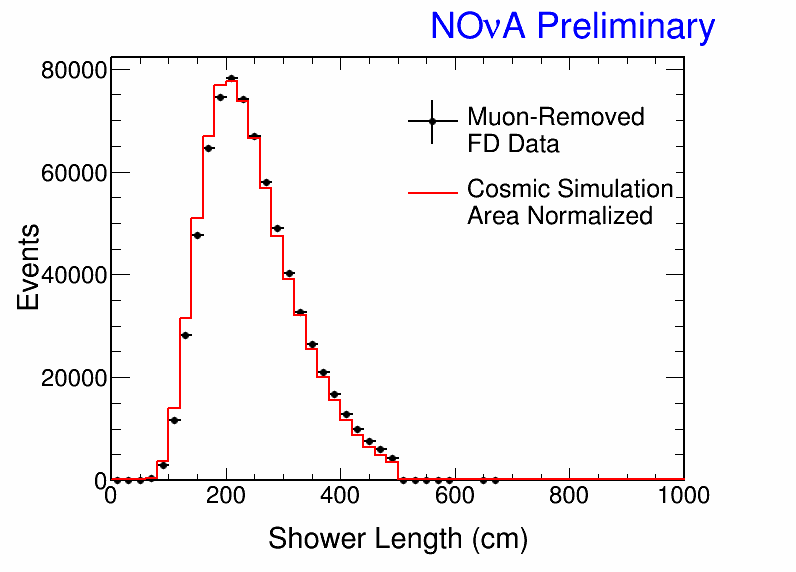}
\includegraphics[width=0.4\textwidth]{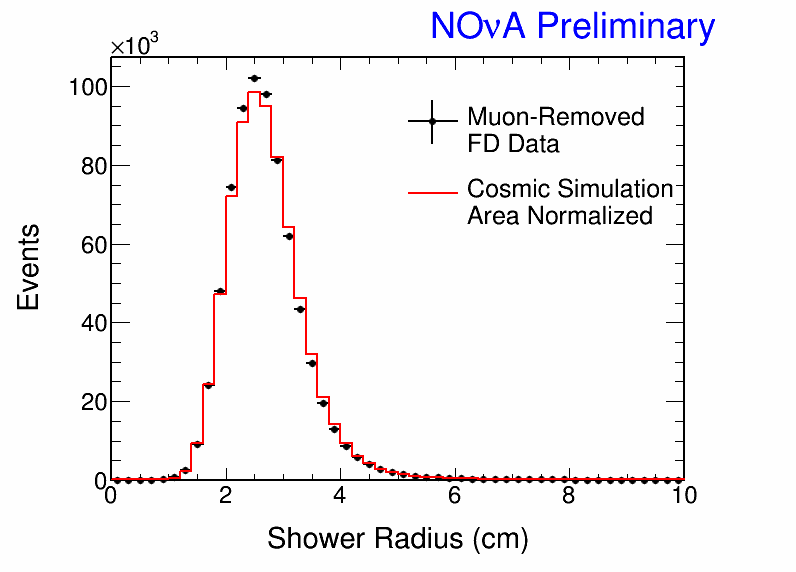}
\includegraphics[width=0.4\textwidth]{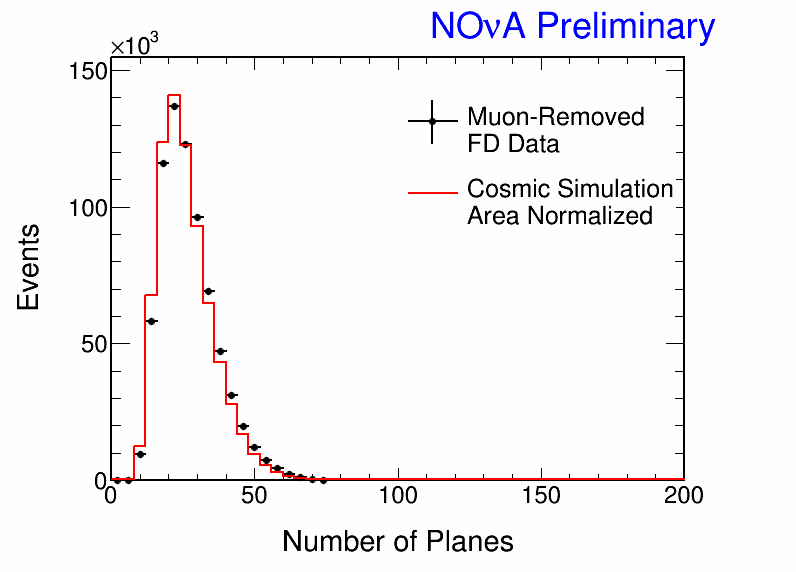}
\includegraphics[width=0.4\textwidth]{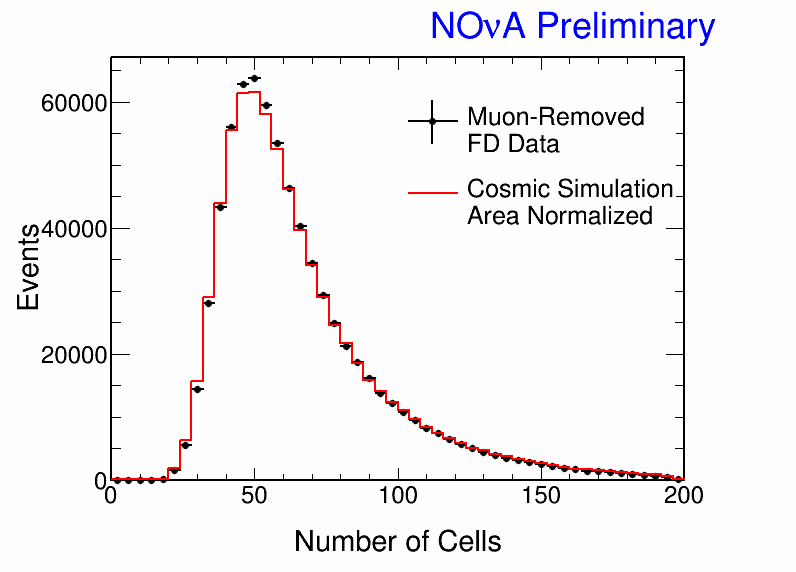}
\end{center}
\caption{The reconstructed shower variables of cosmic EM showers in data and MC, including the reconstructed shower length (top left), width (top right), number of planes (bottom left), and number of cell hits (bottom right).  \label{length}}
\end{figure}

Reconstructed energy and angle are compared between data and MC, and to $\nu_e$ MC events (figure \ref{eshw}). 
Although different in shape, cosmic EM showers cover the same range as $\nu_e$ events. 
Other reconstructed variables such as Shower length, shower width, number of cells, number of planes are also compared between data and MC (figures \ref{length}).  Good data and MC agreement is seen.

The reconstructed prongs are then subject to the particle identification (PID) algorithms. NO$\nu$A has two PID algorithms to distinguish $\nu_e$ signal from background, dominated by $\nu_{\mu}$-NC. The primary PID, Likelihood Based $\nu_e$ Identifier (LID), uses the dE/dx information of a particle to compute the likelihoods that the candidate particle is an electron \cite{lid}\cite{evan}. The alternative algorithm, Library Event Matching (LEM), identifies event types by comparing an unknown trial event to a library of known event from MC \cite{lem}. The distribution shows consistent result between data and MC for both PIDs. While LID identify most of the cosmic EM showers as signal-like (LID$>$0.7), LEM shows most of showers in its background region (LEM$<$0.6). This is due to the fact that LEM is more sensitive to the angle of the showers with respect to the beam direction. A re-weight of cosmic EM showers according to $\nu_e$ events in energy and angle is able to fix this problem.

\begin{figure}
\begin{center}
\includegraphics[width=0.4\textwidth]{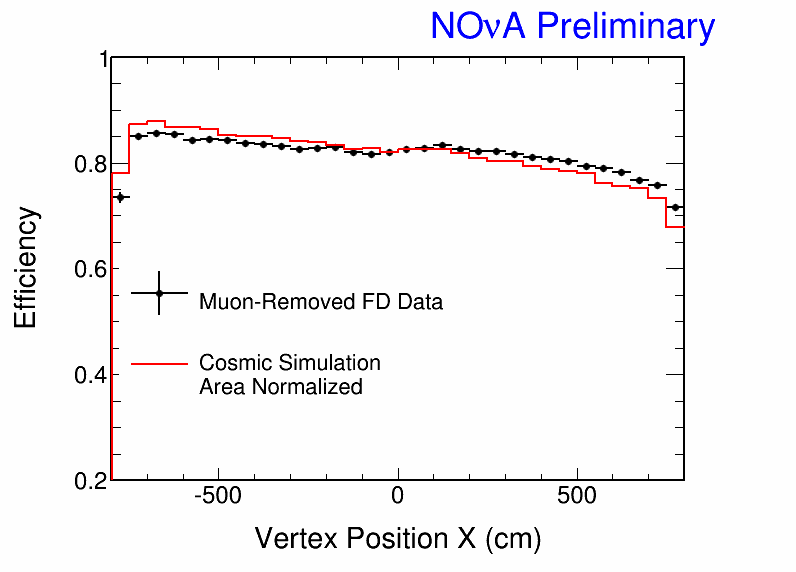}
\includegraphics[width=0.4\textwidth]{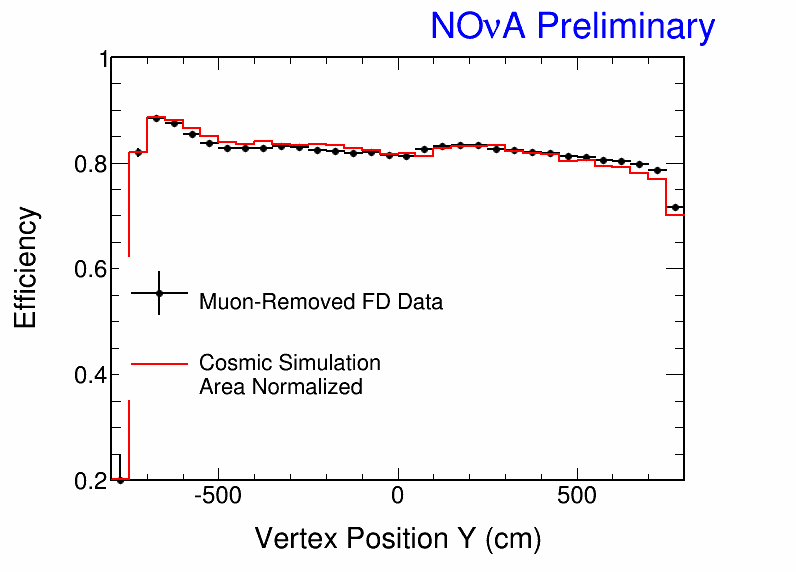}\\
\includegraphics[width=0.4\textwidth]{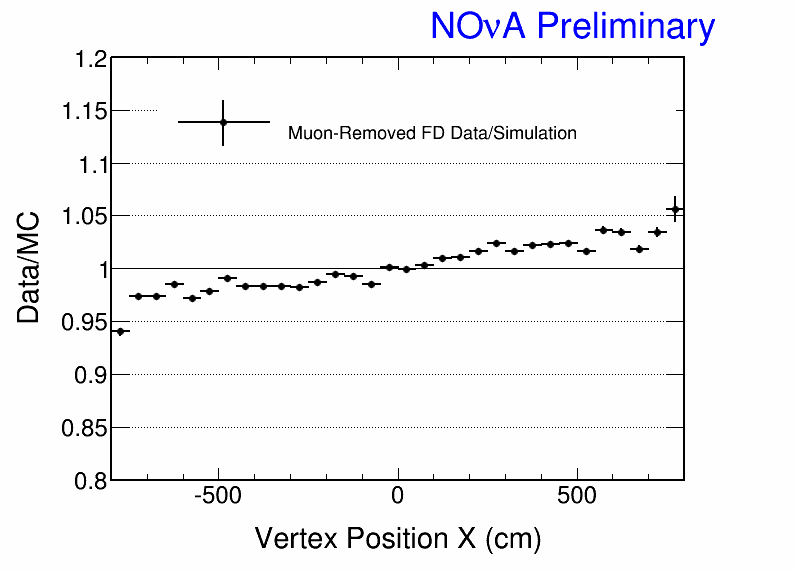}
\includegraphics[width=0.4\textwidth]{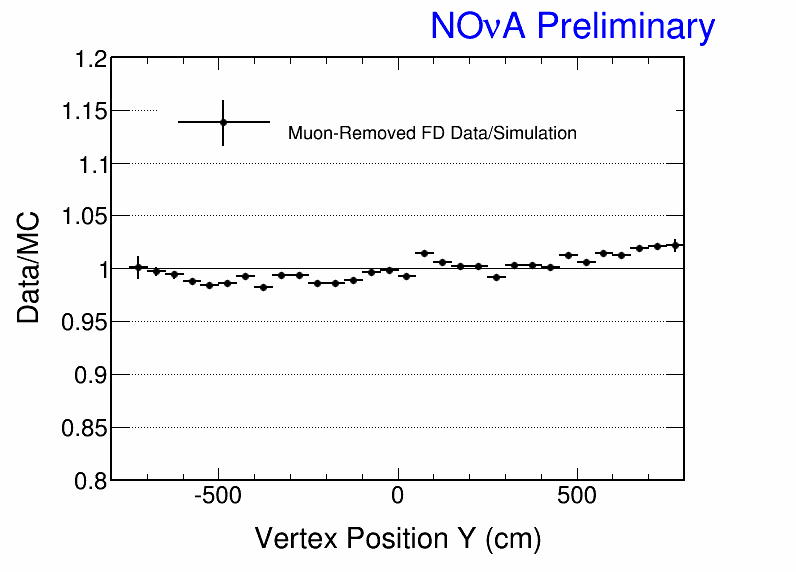}
\end{center}
\caption{Selection efficiency as a function of reconstructed vertex position x (left) and y (right) with LID $>$ 0.7 cut. The bottom plots show the ratio between data and MC efficiencies. The agreement between data and MC is well with in 5 \%.  \label{eff}}
\end{figure}

Efficiencies are calculated as number of showers passing PID cuts (LID$>$0.7) divided by all showers selected as function o vertex position in x and y to check the calibration effects (Figure \ref{eff}). Overall we see a relatively flat efficiency distribution across the detector, with good data and MC agreement. The extent to which the efficiencies from data and MC do not agree motivates a systematic error on the predicted electron neutrino signal efficiency for the first electron neutrino appearance analysis.

\begin{figure}
\begin{center}
\includegraphics[width=0.4\textwidth]{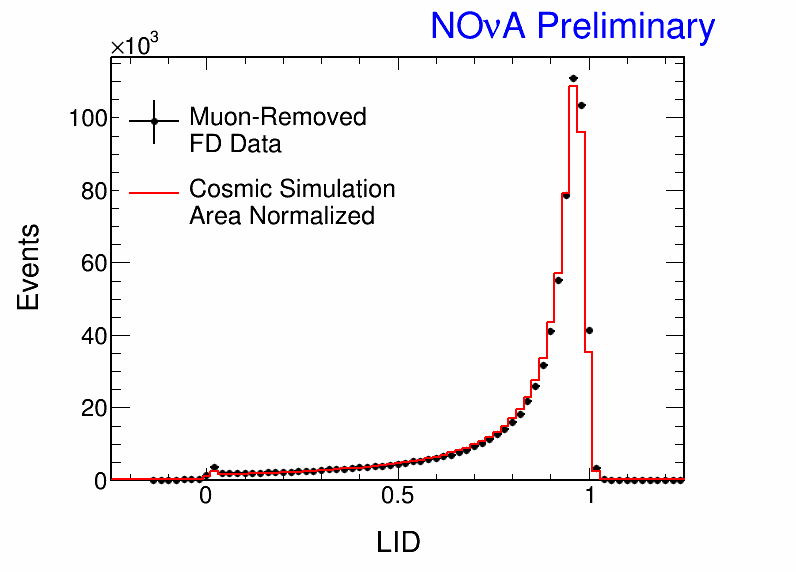}
\includegraphics[width=0.4\textwidth]{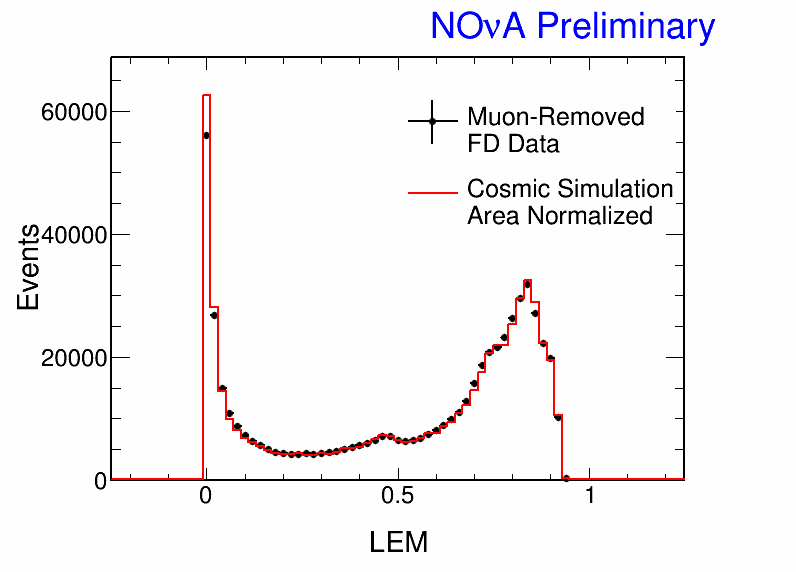}
\end{center}
\caption{LID (left) and LEM (right) distribuion of cosmic EM showers in data and MC. \label{pid}}
\end{figure}

\section{Summary}
We use Cosmic Muon-Removal algorithm for finding and isolating EM showers from data and MC in the NO$\nu$A far detector. The shower digits are put into standard NO$\nu$A $\nu_e$ reconstruction and PID algorithms. A comparison using first analysis data and MC dataset shows consistent distributions, validating the EM shower modeling and reconstruction.
The PID efficiencies as functions of vertex position are calculated with good data/MC agreement. It shows the calibration effects are well controlled.

\end{document}

%% file: econfmacros.tex
\def\beq{\begin{equation}}
\def\eeq#1{\label{#1}\end{equation}}
\def\eeqn{\end{equation}}

\def\beqa{\begin{eqnarray}}
\def\eeqa#1{\label{#1}\end{eqnarray}}
\def\eeqan{\end{eqnarray}}

\let\bar=\overbar

\def\Dslash{\not{\hbox{\kern-4pt $D$}}}
\def\dslash{\not{\hbox{\kern-2pt $\del$}}}

\def\msb{{\bar{\ssstyle M \kern -1pt S}}}




%% file: cosmicmr.bbl
\begin{thebibliography}{99}

\bibitem{novadesign}
D. S. Ayres \textit{et al.} (NO$\nu$A), FERMILAB-DESIGN-2007-01 (2007); R. B. Patterson,
for NO$\nu$A, Nucl. Phys. Proc. Suppl. 235-236, 151 (2013)

\bibitem{numi}
NuMI Technical Design Handbook, NuMI Technical Design Handbook,
http://www-numi.fnal.gov/numwork/tdh/tdh index.html

\bibitem{novafa}
P. Adamson, \textit{et al.} [The NO$\nu$A Collaboration], ``First measurement of electronneutrino
appearance in NO$\nu$A'', FERMILAB-PUB-15-262-ND, to be submitted to Phys. Rev. Lett.

\bibitem{mrcc}
K. Sachdev, ``A Data-Driven Method of Background Prediction at NO$\nu$A'' [arXiv:1501.00968].

\bibitem{lid}
J. Bian, ``First Results of $\nu_e$ Appearance Analysis and Electron Neutrino Identification at NO$\nu$A'', [arXiv:1510.05708]

\bibitem{evan}
E. D. Niner, ``Observation of Electron Neutrino Appearance in the NuMI Beam
with the NO$\nu$A Experiment'',  Ph.D. Thesis, FERMILAB-THESIS-2015-16.

\bibitem{lem}
 C. Backhouse and R. B. Patterson, ``Library Event Matching event classification
algorithm for electron neutrino interactions in the NO$\nu$A detectors'', Nucl.
Instrum. Meth. A 778, \textbf{31} (2015) [arXiv:1501.00968 [physics.ins-det]].

\end{thebibliography}
